\documentclass[aps,twocolumn,prb,showpacs]{revtex4}

\usepackage{amsmath}
\usepackage{bm}
\usepackage{graphicx}

\begin{document}

\title{Intraband exciton relaxation in a biased lattice with long-range
correlated disorder}

\author{E.\ D\'{\i}az}
\author{F.\ Dom\'{\i}nguez-Adame}
\affiliation{GISC, Departamento de F\'{\i}sica de Materiales, Universidad
Complutense, E-28040 Madrid, Spain}

\begin{abstract}

We study numerically the intraband exciton relaxation in a one-dimensional
lattice with scale-free disorder, in the presence of a linear bias. Exciton
transport is considered as incoherent hoppings over the eigenstates of the
static lattice. The site potential of the unbiased lattice is
long-range-correlated with a power-law spectral density $S(k) \sim
1/k^{\alpha}$, $\alpha > 0$. The lattice supports a phase of extended states at
the center of the band, provided $\alpha$ is larger than a critical value
$\alpha_c$ [F.\ A.\ B.\ F.\ de Moura and M.\ L.\ Lyra, Phys. Rev. Lett.
\textbf{81}, 3735 (1998)]. When the bias is applied, the absorption spectrum
displays clear signatures of the Wannier-Stark ladder [E.\ D\'{\i}az \emph{et
al.}, Phys.\ Rev.\ B\textbf{73}, 172410 (2006)]. We demonstrate that in
unbiased lattices and in weakly correlated potentials the decay law is
non-exponential. However, the decay is purely exponential when the bias
increases and $\alpha$ is large. We relate this exponential decay to the
occurrence of the Wannier-Stark ladder in the exciton band.

\end{abstract}


\maketitle

\section{Introduction}

\label{intro}

Long-range correlations with spectral density of the form $S(k) \sim
1/k^{\alpha}$ often arise in nature.~\cite{Paczuski96,Havlin99}  It has been
argued that long-range correlations in nucleotide sequences can explain the
long-distance charge transport in
DNA.~\cite{Carpena02,Yamada04,Albuquerque05,Roche04} This type of correlations
may result in a phase of extended states at the center of the band, provided
$\alpha$ is larger than a critical value $\alpha_c$.~\cite{Moura98} Therefore, a
localization-delocalization transition at the two mobility edges, separating the
phase of extended states at the band center from the localized ones at the band
tails, is found in these systems.~\cite{Izrailev99,Zhang02} It was demonstrated
later that $\alpha_c = 2$ is the universal critical value for the
localization-delocalization transition to occur in the model introduced in
Ref.~\onlinecite{Moura98}, independently of the magnitude of
disorder.~\cite{Shima04}

Bloch-like oscillations~\cite{Bloch28} of quasiparticles are known to arise in
biased one-dimensional (1D) systems with $S(k) \sim 1/k^{\alpha}$ spectral
density.~\cite{Adame03} These oscillations provide a way to measure the energy
width of the delocalized phase since it determines the amplitude of the
oscillations. In addition, the frequency-domain counterpart of the Bloch
oscillations, the so-called Wannier-Stark ladder (WSL),~\cite{Wannier60} was
also observed in the optical absorption spectrum, in spite of the underlying
randomness.~\cite{Diaz06} Strong correlations in the disorder facilitate the
observation of the WSL. At $\alpha > \alpha_c$, when the phase of extended
states emerges at the center of the band, a periodic series of identical peaks
is found to build up at the center of the absorption band. This periodic pattern
was related to the Wannier-Stark quantization of the energy spectrum in the
disordered lattice.~\cite{Diaz06}

In this work we report further progress along this line. We present a model
Hamiltonian for excitons interacting with vibrations of the  host medium. We
assume that there exists an intrinsic bias affecting the exciton dynamics, as
occurs in dendrimers for instance.~\cite{Heijs04} Furthermore, site energies are
assumed to be long-range correlated with $S(k) \sim 1/k^{\alpha}$ spectral
density. The aim of this work is twofold.  First, we analyze how intraband
relaxation, due to the coupling to the vibrations of the glassy host medium,
affects the time behavior of the exciton decay in the \emph{biased\/} lattice.
Second, we look for signatures of the  WSL in the radiative decay of excitons.

The outline of the paper is as follows. In the next section we present our
model, which is based on a tight-binding Hamiltonian of an exciton in a
long-range-correlated potential landscape and subjected to a linear bias. In
Sec.~\ref{Pauli} we recall the basic physics of the exciton intraband relaxation
due to the coupling to host vibrations. Exciton dynamics will be described by a
Pauli master equation for the populations of the exciton
states.~\cite{Bednarz02} The central part of the paper are Secs.~\ref{unbiased}
and~\ref{biased}, where we present the results of numerical simulations of the
time decay of fluorescence after broadband pulse excitation in
disorder-correlated systems. We discuss in detail its dependence on the driving
parameters of the model (bias magnitude, disorder strength and correlation
exponent $\alpha$) and provide an evidence of that the WSL reveals in the 
intraband exciton relaxation. We find that the fluorescence decay exponentially
when localization by the bias prevails over localization by disorder. We relate
this exponential decay to the occurrence of the WSL. When disorder is large
enough or long-range correlations are weak ($\alpha<\alpha_c$), the time
dependence of the fluorescence decay is non-exponential. Finite size effects on
the time decay of fluorescence are analyzed in Sec.~\ref{finite-size}. Finally,
Sec.~\ref{conclusions} concludes the paper.

\section{Model Hamiltonian}
\label{model}

We consider a biased tight-binding model with diagonal disorder on an otherwise
regular 1D open lattice of spacing unity and $N$ sites ($N$ is assumed to be
even). We assign two levels to each lattice site, ground and excited, and
consider optical transitions between them with transition energy ${\mathcal
E}_n=\bar{\mathcal E} + \varepsilon_n$. The stochastic part $\varepsilon_n$ is
generated according to~\cite{Moura98}
\begin{subequations}
\label{disorder}
\begin{equation}
\varepsilon_{n} =\sigma C_\alpha \sum_{k=1}^{N/2}
\frac{1}{k^{\alpha/2}}\, \cos\left(\frac{2\pi kn}{N} + \phi_k\right)\ ,
\end{equation}
where the normalization constant is given by
\begin{equation}
C_\alpha = \sqrt{2}\left(\sum_{k=1}^{N/2} \frac{1}{k^{\alpha}}\right)^{-1/2} \ .
\end{equation}
\end{subequations}
Here  $\phi_1,\ldots,\phi_{N/2}$ are uncorrelated random phases uniformly
distributed within the interval $[0,2\pi]$. The distribution~(\ref{disorder})
has zero mean $\langle \varepsilon_{n}\rangle = 0$ and standard deviation
$\langle \varepsilon_{n}^2\rangle^{1/2} = \sigma$, where $\langle \ldots
\rangle$ indicates averaging over realizations of random phases $\phi_k$. The
quantity $\sigma$ will be referred to as magnitude of disorder.

The model Hamiltonian is
\begin{eqnarray}
    {\cal H} & = & \sum_{n=1}^{N} \left[{\mathcal E}_n
    - U\left(n - \frac{N}{2}\right) \right]\,|n\rangle\langle n| \nonumber\\
    & - & J\sum_{n=1}^{N-1}\Big(|n\rangle\langle n+1|
    + |n+1\rangle\langle n|\Big)\ .
\label{hamiltonian}
\end{eqnarray}
Here, $|n\rangle$ denotes the state in which the $n\,$th site is excited,
whereas all the other sites are in the ground state.  The term $-U(n - N/2)$
describes the linear bias.The intersite transfer integrals
in~(\ref{hamiltonian}) are restricted to nearest-neighbors, and it is set to
$-J$ over the entire lattice. Also, we set $\bar{\mathcal E} = 0$ hereafter
without loss of generality.

As we already mentioned in Sec.~\ref{intro}, the unbiased model supports a phase
of extended states at the center of the band, provided the correlation exponent
$\alpha$ is larger than a critical value $\alpha_c$. On the contrary, all the
states are localized at $\alpha < \alpha_c$.

\section{Intraband relaxation}
\label{Pauli}

Excitons can hop between the eigenstates of the Hamiltonian~(\ref{hamiltonian})
only if coupling to vibrations is taken into account. We assume that this
coupling is weak and do not consider polaron effects. The exciton-vibration
interaction causes the \emph{incoherent\/} hopping of excitons from one
eigenstate to another. Only one-phonon processes are considered throughout the
paper. We take the transition rate from the state $\psi_\mu$ (with  energy
$E_\mu$) to the state $\psi_\nu$ (with energy $E_\nu$) according to (see
Ref.~\onlinecite{Bednarz02} for further details)
\begin{equation}
W_{\mu\nu}=W_0\,S(|\Delta E_{\mu\nu}|)\,F(\Delta E_{\mu\nu},T)\,
\mathcal{I}_{\mu\nu}\ ,
\label{1Wkk}
\end{equation}
with $\Delta E_{\mu\nu}\equiv E_\mu-E_\nu$. The constant $W_0$ characterizes the
amplitude of transitions.  A spectral density function of the form $S(|\Delta
E|)=|\Delta E|/J$ holds for glassy  hosts.~\cite{Shimizu01,Bednarz01}
Temperature $T$ enters into this expression through the function $F(\Delta
E,T)=\theta(-\Delta E)+n(\Delta E,T)$, where $\theta$ is the Heaviside step
function and  $n(\Delta E,T) = [\,\exp(\Delta E/k_BT) - 1]^{-1}$ is the
occupation number of the vibration mode with frequency $\Delta E/\hbar$. The
parameter
\begin{equation}
\mathcal{I}_{\mu\nu} \equiv \sum_{n=1}^N \psi_{\mu,n}^2\psi_{\nu,n}^2
\label{overlap}
\end{equation}
represents the overlap integral of exciton probabilities for the states
$\psi_\mu$ and $\psi_\nu$.  Notice that the transition rates meet the principle
of detailed balance: $W_{\nu\mu} = W_{\mu\nu}\exp(\Delta E_{\mu\nu}/k_BT)$. 

We describe the process of exciton relaxation by means of the Pauli master
equation for the level population $P_\mu$ of the $\mu$\,th exciton eigenstate
\begin{equation}
\frac{dP_\mu}{dt} = -\gamma_\mu P_\mu + \sum_{\nu = 1}^N
(W_{\mu\nu}\,P_{\nu} - W_{\nu\mu}\,P_\mu)\ , 
\label{P_mu}
\end{equation}
where $\gamma_\nu = \gamma\,f_\mu$ is the spontaneous emission rate of the
$\mu$\,th exciton state, while $\gamma$ is that of a monomer,
$f_\mu=(\sum_{n=1}^N \psi_{\mu,n})^2$ being the dimensionless oscillator
strength.  After broadband pulse excitation, each level is populated according
to its oscillator strength, namely $P_\mu(0)=f_\mu/N$. Consequently, the initial
total population is normalized to unity, $\sum_\mu P_\mu(0) = 1$. The normalized
fluorescence after pulse excitation is obtained from
\begin{equation}
I(t)=N\,\frac{\sum_{\mu=1}^{N}f_{\mu}P_{\mu}(t)}{\sum_{\mu=1}^{N}f_{\mu}^2}\ ,
\label{fluorescence}
\end{equation}
which is the magnitude of interest in this work.

\section{Unbiased system} 
\label{unbiased}

\begin{figure}[ht]
\centering
\includegraphics[width=80mm,clip]{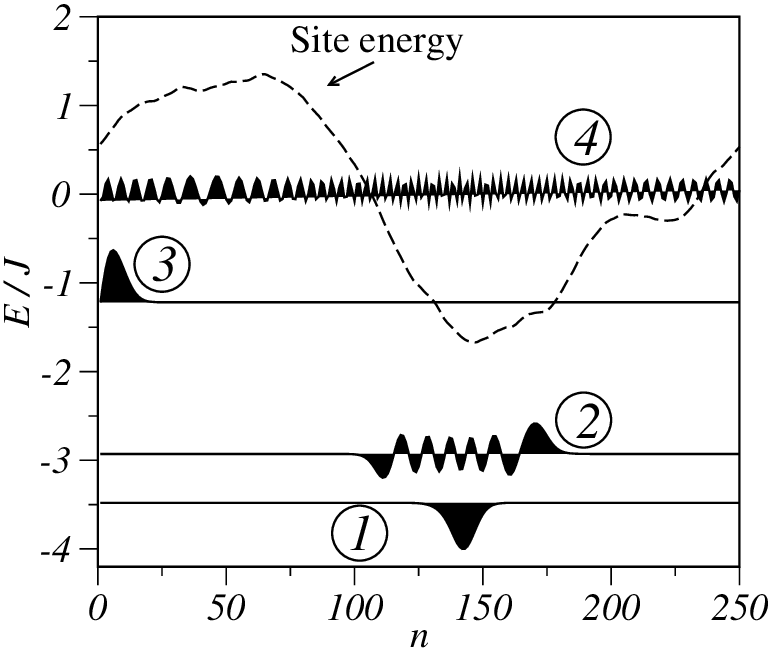}
\caption{A subset of eigenstates for a typical realization of the random energy
potential (dashed line), for a lattice of size $N=250$, magnitude of disorder
$\sigma = J$, and correlation exponent $\alpha=4$ (larger than the critical
value $\alpha_c = 2$). The baseline indicates the energy of the eigenstate. The
states $1$ and $3$ are those which larger oscillator strength.}
\label{fig01}
\end{figure}

We first discuss the exciton decay of the unbiased system ($U =0$), aiming to
separate the effects of bias from those related to the disordered nature of the
model. We have numerically diagonalized the Hamiltonian~(\ref{hamiltonian}) by
means of standard methods and obtained the eigenstates $\psi_\mu$ and
eigenvalues $E_\mu$ for different values of the correlation exponent $\alpha$
and magnitude of disorder $\sigma$, considering open linear lattices. In
Fig.~\ref{fig01} we show a subset of eigenstates obtained for a typical random
realization of the potential landscape for $\alpha=4$, larger than the critical
value $\alpha_c = 2$. The baselines display the corresponding eigenenergies. The
lowest state in Fig.~\ref{fig01} (labeled $1$) shows a bell-like shape and
carries a large oscillator strength. There are several states of such type (not
shown in Fig.~\ref{fig01}), which are close in energy to the state $1$ and do
not overlap with each other. On increasing the energy, one observes eigenstates,
like the one labeled $2$ in Fig.~\ref{fig01}, which are more extended, as
compared to the lowest one, and present several nodes within the localization
segment. The oscillator strength of such states is much smaller  (dark states).
Remarkably, going further up in energy we again find bell-like states, as the
one labeled $3$ in Fig.~\ref{fig01}, which are characterized by a large
oscillator strength. Finally, on approaching the center of the band one expects
the occurrence of extended states. The state $4$ in Fig.~\ref{fig01} represents
an example. 

The picture presented above is different when $\alpha$ is smaller than the
critical value $\alpha_c = 2$. The level structure is then similar to what it is
found in the standard 1D Anderson model (uncorrelated disorder). Bell-like
states lay at the bottom of the band and dark states, localized over larger
segments, are higher in energy (see, e.g., Fig.~1 in
Ref.~\onlinecite{Malyshev03}). In addition, there are not extended states like
that labeled $4$ in Fig.~\ref{fig01} at $\alpha<\alpha_c$.

Having in mind the different level structure for $\alpha$ larger and smaller
that the critical value $\alpha_c$, we obtained the normalized fluorescence
after pulse excitation~(\ref{fluorescence}). Hereafter we restrict ourselves to
$T=0$; we have checked that our main conclusions are valid when the temperature
is a fraction of $J/k_B$. This can be understood from the large energy 
difference between states $1$ and $3$ in Fig.~\ref{fig01}, which are those 
strongly coupled to the light due to their high oscillator strength. In other
words, it is highly improbable that an initial excitation of the state $1$ can
be transferred to the state $3$ after several hops, at least at low $T$. Also,
we choose the parameter $W_0=\gamma$ hereafter, so that radiative decay is
favored with respect to intraband relaxation. This statement can be understood
from previous estimations in unbiased lattices with uncorrelated
disorder.~\cite{Malyshev03-bis} When $\sigma=J$, the largest intraband
transition rate $W_{\mu\nu}$ is of the order $\sim 0.06W_0$ but the higher
spontaneous emission rate $\gamma_\mu$ is about $\sim 9\gamma$. 

\begin{figure}[ht]
\centering
\includegraphics[width=80mm,clip]{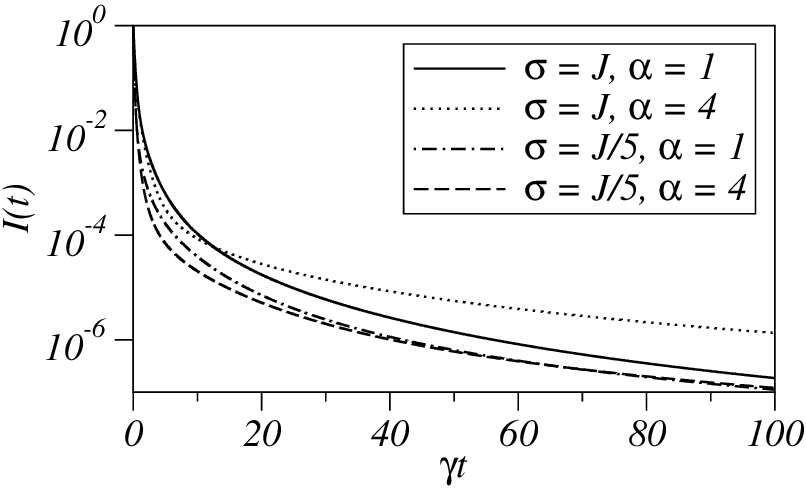}
\caption{Time dependence of the fluorescence decay after pulse excitation in
weakly ($\alpha=1$) and strongly ($\alpha=4$) correlated 1D lattices when
$U=0$, for two different magnitudes of disorder. Lattice size is $N=250$ and the
results comprise the average over $200$ realizations of the disorder.}
\label{fig02}
\end{figure}

Figure~\ref{fig02} shows the fluorescence decay after pulse excitation as a 
function of $\gamma t$ for various set parameters and $W_0=\gamma$. The
lattice size is $N=250$ and the results comprises the average over $200$
realizations of the disorder. In all cases the decay is clearly non-exponential.
At large times the decay law fits well a stretched exponential $I(t) \sim
\exp\left[-(t/\tau_{\infty})^{\beta}\right]$, where both $\tau_\infty$ and
$\beta<1$ depend on the correlation exponent $\alpha$. The fluorescence curves
are similar when the magnitude of disorder $\sigma$ is small, as can be seen in
Fig.~\ref{fig02}. However, on increasing the magnitude of disorder the
fluorescence decay is slower in strongly correlated systems. 

The slowing down of the fluorescence decay can be understood as follows. When
$\alpha$ is smaller than $\alpha_c$, high energy states are weakly coupled to
the light due to their low oscillator strength. The excitation is then
transferred from the high energy states (dark states) to those at the bottom of
the band (bell-like states). Being stuck in the bottom state, the exciton emits
a photon on average after a time $1/\gamma_1$ has elapsed. On the contrary, when
$\alpha$ is larger than $\alpha_c$, the initial exciton population does not
decrease monotonously on increasing energy due to occurrence of bell-like states
like that labeled $3$ in Fig.~\ref{fig01} (recall that the initial level
population is proportional to the corresponding oscillator strength). In
contrast to bell-like states at the bottom of the band, those highly populated
states after excitation may not decay radiatively: an exciton initially located
at state $3$ in Fig.~\ref{fig01} may be transferred to a state of lower energy
whose oscillator strength is vanishingly small (e.g. state $2$ is
Fig.~\ref{fig01}). The exciton is then scattered  to bottom states (e.g. state
$1$ is Fig.~\ref{fig01}) and then emits a photon. Consequently, on average there
are more intraband scattering events through dark states when $\alpha$ is large,
yielding a slowdown observed in Fig.~\ref{fig02}.

\section{Biased system} 
\label{biased}

In disorder-free systems, switching the bias on results in a reorganization of
the level structure, which becomes equally spaced with level spacing
$U$.~\cite{Wannier60} The corresponding eigenstates become localized  within a
localization segment of length $L_U=W/U$ in units of the spatial period, where
$W$ is the bandwidth. In 1D disorder-free systems this bandwidth is $W=4J$. This
structure is revealed in photoluminescence~\cite{Mendez88,Agullo89} and
photoconductivity~\cite{Saker91} spectra as a series of equally spaced peaks.
Deviations from perfect periodicity broadens the peaks and makes them unresolved
when disorder is uncorrelated and large. As mentioned in the Introduction,
long-range correlations facilitate the occurrence of the WSL even if disorder is
large.~\cite{Diaz06} In this case, a periodic pattern is found to build up at
the center of the optical absorption band, where the absorption spectrum
lineshape is obtained from
\begin{equation}
A(E)= \frac{1}{N}\Bigg\langle \sum_{\mu=1}^{N}
f_\mu \delta(E - E_{\mu})\Bigg\rangle \ .
\label{lineshape}
\end{equation}
The period of the periodic pattern is equal to $U$, as for the WSL in an ideal
lattice, and independent of the system size $N$. This pattern was attributed to
the Wannier-Stark quantization of the energy spectrum in the long-range
correlated disordered lattice.~\cite{Diaz06}

\begin{figure}[ht]
\centerline{\includegraphics[width=80mm,clip]{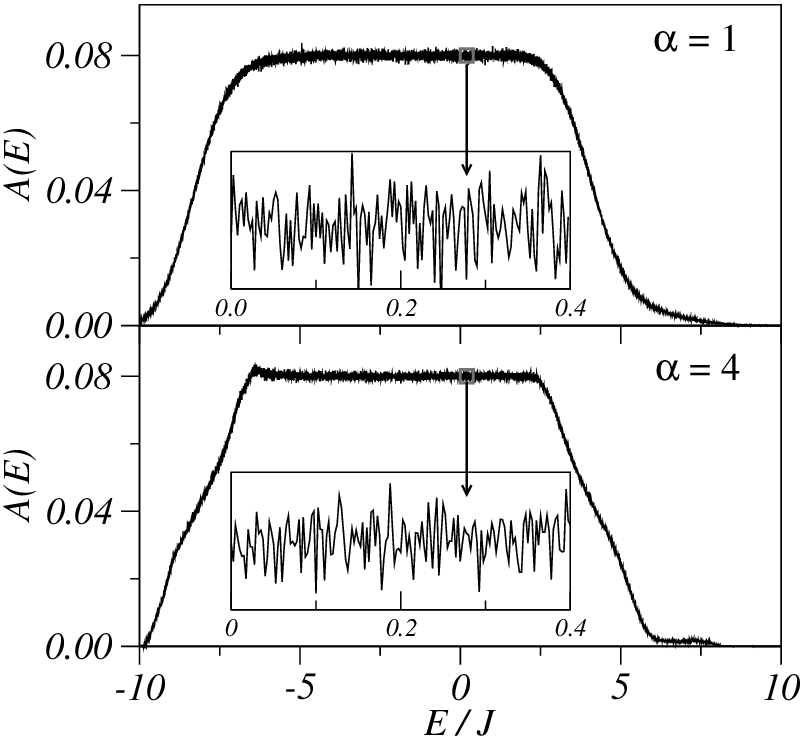}}
\caption{ Absorption spectra of biased chains ($U = 0.05J$) with $N = 250$ sites
calculated for two values of the correlation exponent $\alpha$, shown in the
plot. The magnitude of disorder is $\sigma = J$. Each curve were obtained after
averaging over $10^6$ realizations of disorder. Insets show enlarged views of
the spectra within the small boxes.} 
\label{fig03}
\end{figure}

\begin{figure}[ht]
\centering
\includegraphics[width=80mm,clip]{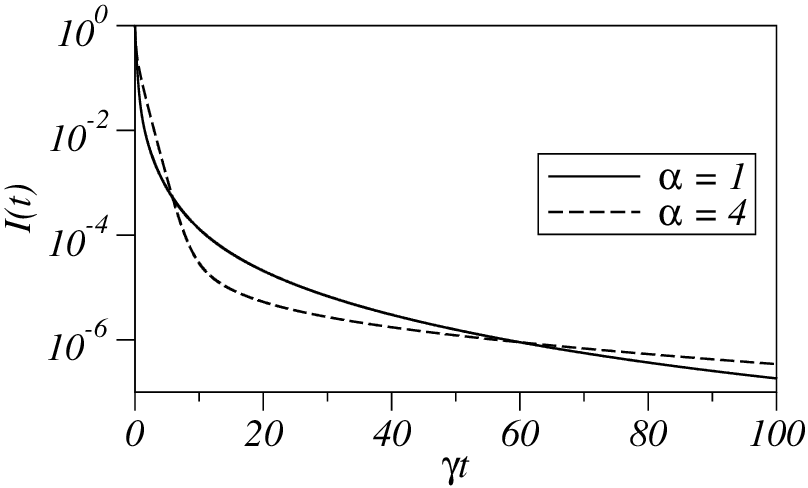}
\caption{Time dependence of the fluorescence decay after pulse excitation in
weakly ($\alpha=1$) and strongly ($\alpha=4$) correlated 1D lattices when 
$U=0.05J$ and $\sigma=J$. Lattice size is $N=250$ and the results comprises the
average over $200$ realizations of the disorder.}
\label{fig04}
\end{figure}

In Fig.~\ref{fig03} we plotted the absorption spectra calculated for disordered
biased chains of $N = 250$ sites, choosing the magnitude of disorder $\sigma =
J$ and the bias magnitude $U = 0.05J$. Averages over $10^6$ realizations of
disorder were performed in Eq.~(\ref{lineshape}). Two values of the correlation
exponents were considered, $\alpha = 1$ and $\alpha = 4$. The spectra broaden as
compared to those of the unbiased systems (not shown in the figure). In
Ref.~\onlinecite{Diaz06} we proved that the spectrum broadens when  $\sigma^{*}
< UN$, namely when the absorption line width $\sigma^{*}$  in the unbiased
system is smaller than the potential energy drop across the entire lattice $UN$.
The width of the absorption spectrum (FWHM) when $U$ is low is of the order of
$UN=12.5$, as it is already observed in Fig.~\ref{fig03}. However, the
absorption  spectra are still featureless and no signatures of the WSL are
detected. The corresponding fluorescence decay is shown in Fig.~\ref{fig04},
where we observe that the decay is non-exponential. At large time the decay is
slower on increasing the magnitude of the correlation, similarly to what was
found in the absence of bias (see Fig.~\ref{fig02}).

\begin{figure}[ht]
\centerline{\includegraphics[width=80mm,clip]{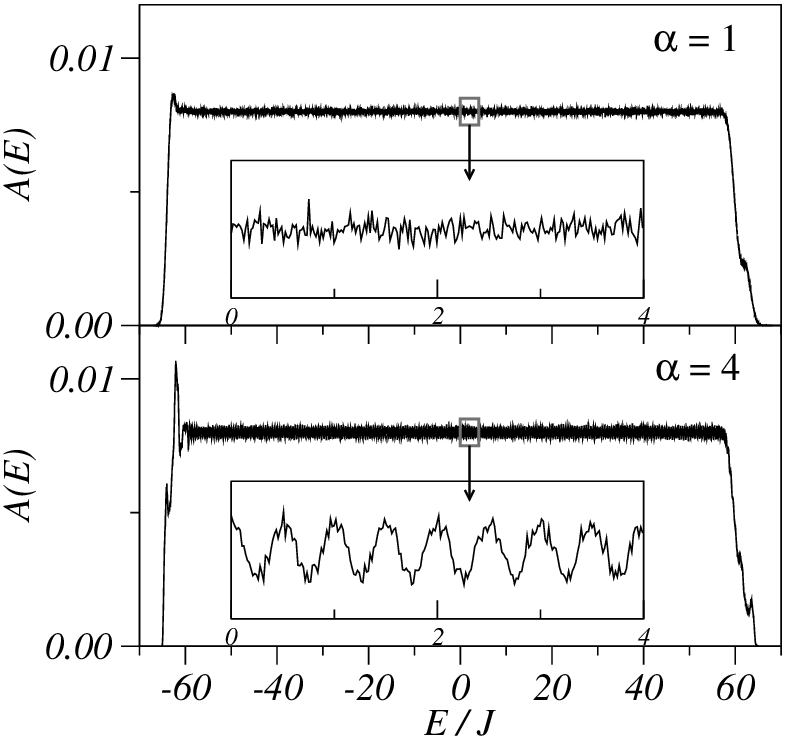}}
\caption{ Absorption spectra of biased chains ($U = 0.5J$) with $N = 250$ sites
calculated for two values of the correlation exponent $\alpha$, shown in the
plot. The magnitude of disorder is $\sigma = J$. Each curve were obtained after
averaging over $10^6$ realizations of disorder. Insets show enlarged views of
the spectra within the small boxes.} 
\label{fig05}
\end{figure}

In Fig.~\ref{fig05} we plotted the absorption spectra calculated for a high
value of the bias, $U=0.5J$. We observe in the upper panel that at $\alpha<
\alpha_c$ the spectrum remains structureless. However, in the lower panel we see
that for strong correlations in disorder, when $\alpha > \alpha_c$, the spectrum
presents a periodic pattern which is not masked by the stochastic disorder
fluctuations (see the inset in the lower panel). Most important, the period of
the modulation is exactly equal to $U = 0.5J$, as results from the occurrence of
the WSL in the energy spectrum of the system.

\begin{figure}[ht]
\centering
\includegraphics[width=80mm,clip]{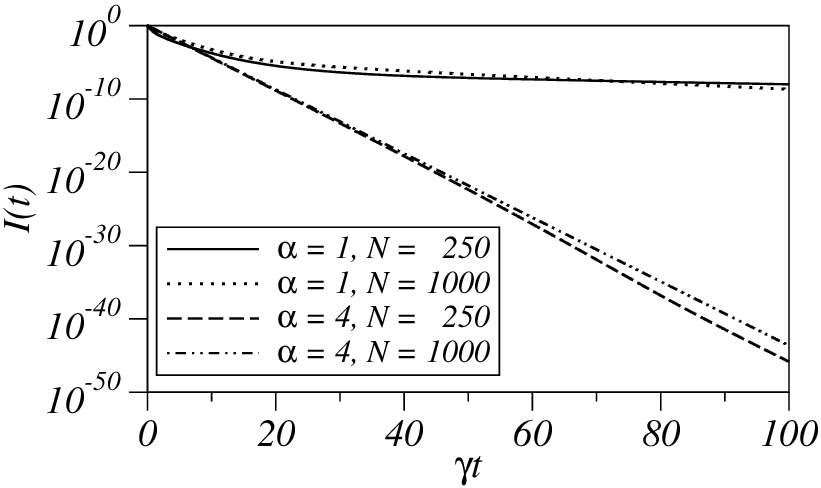}
\caption{Time dependence of the fluorescence decay after pulse excitation in
weakly ($\alpha=1$) and strongly ($\alpha=4$) correlated 1D lattices when 
$U=0.5J$ and $\sigma=J$. Lattice size is $N=250$ and the results comprises the
average over $200$ realizations of the disorder. For comparison, results for
larger systems ($N=1000$) are also shown, comprising averages over $50$
realizations of the disorder.}
\label{fig06}
\end{figure}

As mentioned above, long-range correlations in disorder strongly affect the
optical response of the system. We now present evidences that they also have
remarkable impact on the fluorescence decay. More important, we will show that
the peculiarities found in the fluorescence decay can be related to the
occurrence of the WSL. To this end, we numerically solve ~(\ref{hamiltonian}) 
and~(\ref{P_mu}) to obtain the time-dependence of the fluorescence when $U\neq
0$. Figure~\ref{fig06} shows the fluorescence decay after pulse excitation as a 
function of $\gamma t$ for various set parameters and $W_0=\gamma$.  The lattice
size is $N=250$, the bias is $U=0.5J$ and the results comprises the average over
$200$ realizations of the disorder. We observe that the fluorescence decay is
still non-exponential when $\alpha<\alpha_c$. However, the decay is exponential
in the opposite limit, when $\alpha>\alpha_c$.  It is to be noticed that the
system size $N$ should be sufficiently large enough to yield the power-law
spectral density $S(k) \propto k^{-\alpha}$. If $N$ is inadequately small,
$S(k)$ would not obey a power law. To ascertain whether results mentioned above
can be attributed to long-range spatial correlations, we also studied larger
system sizes. Figure~\ref{fig06} also shows the fluorescence decay when
$N=1000$, the results comprising the average over $50$ realizations of the
disorder. The  non-exponential and exponential decays for weakly and strongly
correlated disorder respectively are clearly observed (see
Sec.~\ref{finite-size} below for a detailed discussion of finite-size
effects).

\begin{figure}[ht]
\centerline{\includegraphics[width=80mm,clip]{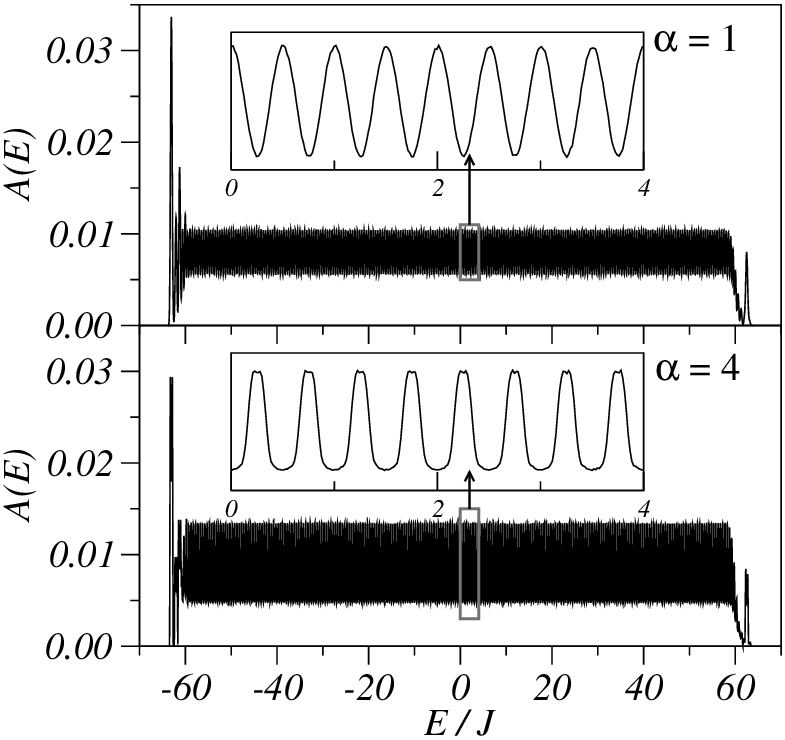}}
\caption{ Absorption spectra of biased chains ($U = 0.5J$) with $N = 250$ sites
calculated for two values of the correlation exponent $\alpha$, shown in the
plot. The magnitude of disorder is $\sigma = 0.2J$. Each curve were obtained
after averaging over $10^6$ realizations of disorder. Insets show enlarged views
of the spectra within the small boxes.} 
\label{fig07}
\end{figure}

\begin{figure}[ht]
\centering
\includegraphics[width=80mm,clip]{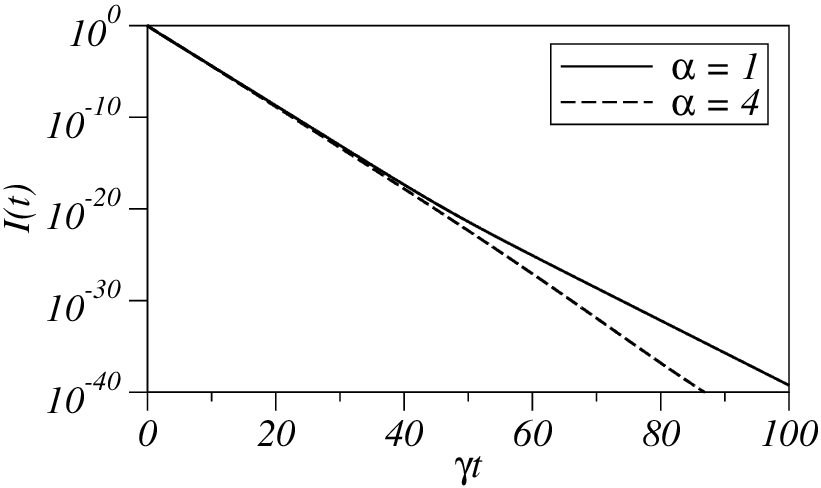}
\caption{Time dependence of the fluorescence decay after pulse excitation in
weakly ($\alpha=1$) and strongly ($\alpha=4$) correlated 1D lattices when 
$U=0.5J$ and $\sigma=0.2J$. Lattice size is $N=250$ and the results comprises
the average over $200$ realizations of the disorder.}
\label{fig08}
\end{figure}

In Ref.~\onlinecite{Diaz06} we claimed that the WSL can be resolved even if
correlations are weak. We pointed out that the condition for the WSL to appear
in the absorption spectrum is $\sigma> \sqrt{UJ}$ in the weakly correlated
case.  We calculated the absorption spectrum lineshape for magnitudes of
disorder $\sigma = 0.2J$ and bias $U = 0.5J$, when this inequality holds. The
results, obtained for $\alpha = 1$ and $\alpha = 4$ are depicted in
Fig.~\ref{fig07} (upper and lower panels, respectively). One observes that the
absorption spectrum shows a resolved WSL structure for both values of $\alpha$.
Figure~\ref{fig08} indicates that the fluorescence decay exponentially with time
when $\alpha>\alpha_c$ and biexponentially when $\alpha<\alpha_c$.

\begin{figure}[ht]
\centering
\includegraphics[width=80mm,clip]{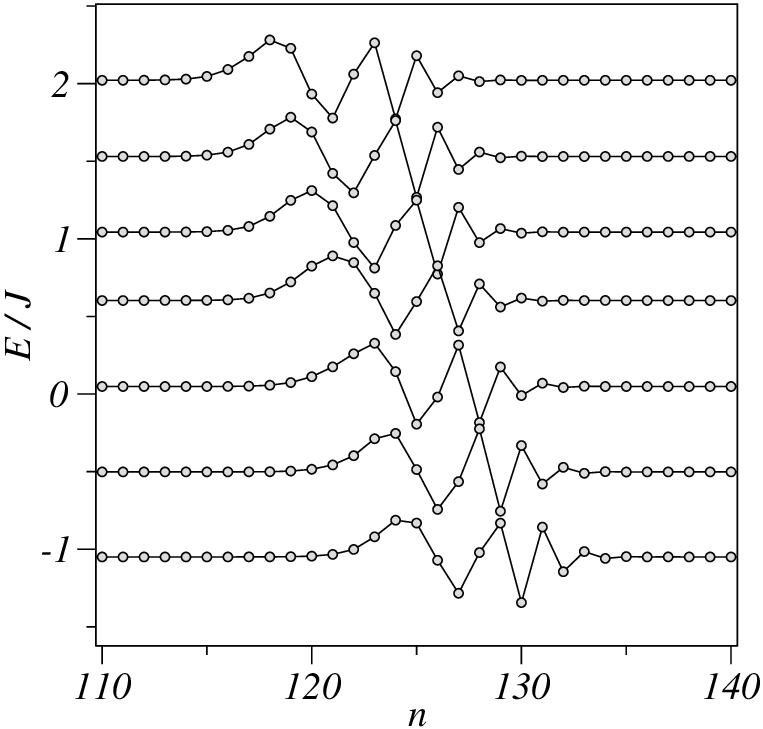}
\caption{A subset of eigenstates for a typical realization of the random energy
potential when $U=0.5$ for a lattice of size $N=250$, magnitude of disorder
$\sigma = 0.2J$, and correlation exponent $\alpha=1$. The baseline indicates the
energy of the eigenstate.}
\label{fig09}
\end{figure}

Results shown in Figs.~\ref{fig06} and~\ref{fig08} provide compelling evidences
that the occurrence of the WSL in the energy spectrum may result in an
exponential decay of the fluorescence. In the upper panel of Fig.~\ref{fig05} we
observed the absence of WSL signatures in the optical absorption spectrum when
$\alpha <\alpha_c$ and the magnitude of disorder $\sigma$ is large.
Simultaneously, the fluorescence decay is non-exponential, as shown in
Fig.~\ref{fig06} and Fig.~\ref{fig08}. This behavior can be understood from the
fact that when $\sigma$ is large and correlations are weak ($\alpha <\alpha_c$)
the oscillator strength depends on the energy state, similarly to what it is
found in the absence of bias (see Sec.~\ref{unbiased}). This finally leads to a
non-exponential decay of the fluorescence. However, when correlations are
strong, $\alpha >\alpha_c$, or the magnitude of disorder $\sigma$ is smaller
than $\sqrt{UJ}$, the occurrence of the WSL ladder is accompanied by an
exponential decay of the fluorescence, as shown if Fig.~\ref{fig06}. In this
case localization by the bias $U$ prevails over localization by disorder. When
the WSL arises, the shape of the eigenfunction is almost the same for all
states, except for a trivial space shift and excluding finite size effects (see
Fig.~\ref{fig09}). Consequently, the oscillator strength of all states has the
same value, yielding a single exponential in the decay law of the fluorescence. 

\section{Finite size effects} 

\label{finite-size}

The peculiarities observed at the low- and high-energy sides of the absorption
spectra shown in Figs.~\ref{fig05} and~\ref{fig07} are associated  with finite
size effects.~\cite{Diaz06} The levels of these regions of the energy spectrum
are formed by the states localized close to the system ends. Because of that,
the corresponding eigenfunctions differ from those at the band center. As a
consequence, not all eigenstates present exactly the same value of the 
oscillator strength and slight deviation from the single exponential decay of
the fluorescence can be detected. 

To understand the relevance of finite size effects, we focus on the biased,
disorder-free lattice. In the thermodynamic limit ($N\to\infty$), the normalized
eigenstates of the Hamiltonian~(\ref{hamiltonian}) with $\sigma=0$ are expressed
in terms of the Bessel functions as $\psi_{\mu,n}= J_{n-\mu}(2J/U)$ (see
Ref.~\onlinecite{Fukuyama73}). The corresponding eigenenergies are $E_\mu=\mu
U$, $\mu$ being an integer (i.e. the WSL structure). The oscillator strength can
also be calculated exactly in this limiting case $f_{\mu}=1$, which takes the
same value for all states, as expected. Remarkably, the  oscillator strength
becomes also independent of the bias $U$ in large disorder-free systems.
From~(\ref{fluorescence}) we conclude that the normalized fluorescence intensity
is nothing but the survival probability $I(t)=\sum_{\mu=1}^{N}P_{\mu}(t)$ in
this case. Summing over all states in~(\ref{P_mu}) we obtain $\dot{I}(t) =
-\gamma I(t)$, leading to $I(t)=\exp(-\gamma t)$. Consequently, the fluorescence
decay in biased disorder-free lattices is the same as in the isolated monomer
(i.e. single exponential with decay time $1/\gamma$), when the system is large
enough.

\begin{figure}[ht]
\centering
\includegraphics[width=80mm,clip]{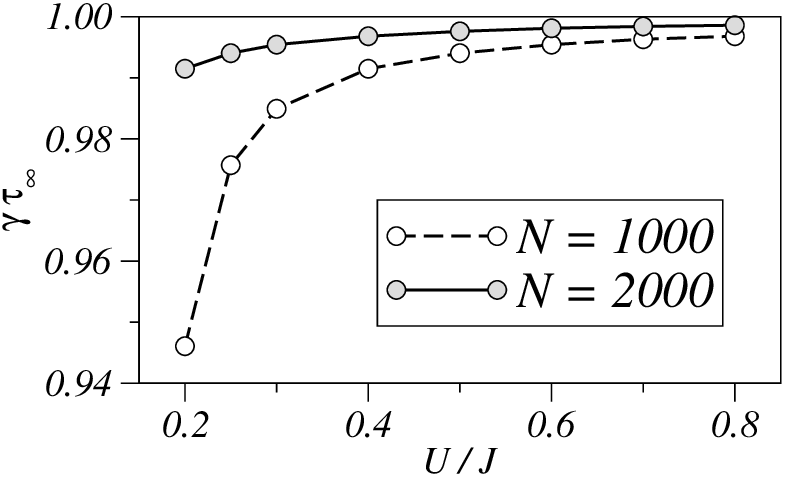}
\caption{Bias dependence of $\gamma\tau_\infty$ for different system sizes,
when $\sigma=0$.}
\label{fig10}
\end{figure}

As mentioned in the preceding paragraph, when $N\to\infty$ the occurrence of the
WSL in the energy spectrum ($E_\mu=\mu U$) in disorder-free systems leads to an
exponential decay of the fluorescence. The decay time is found to be the same as
in the single monomer and consequently it is independent of the bias. We have
performed simulations in finite system when $\sigma=0$ to obtain the decay time
$\tau_\infty$ from the fluorescence intensity curves when $t\to\infty$. Results
are shown in Fig.~\ref{fig10}, where we observe that $\gamma\tau_\infty$
approaches the theoretical limit ($\gamma\tau_\infty=1$) on increasing both the
system size or the bias. The localization length $L_U$ due to the presence of
the bias decreases on increasing $U$, leading to smaller finite size effects.

\begin{figure}[ht]
\centering
\includegraphics[width=80mm,clip]{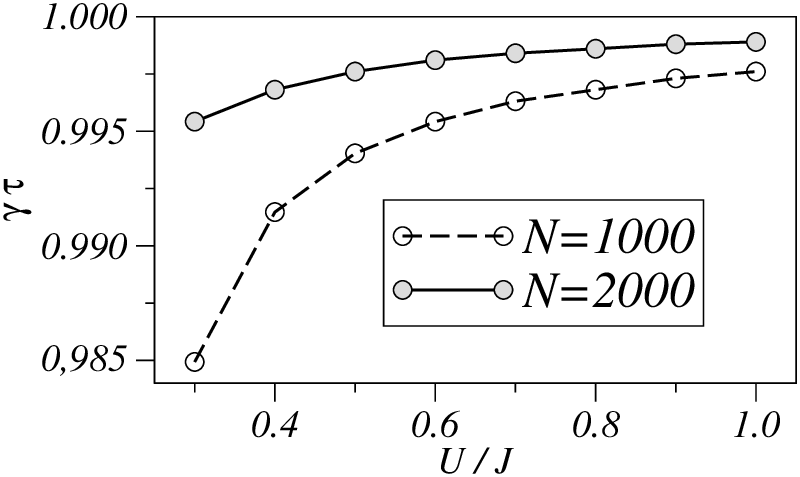}
\caption{Bias dependence of $\gamma\tau_\infty$ for different system sizes, when
$\sigma=J$ and $\alpha=4$. Results comprise averages over $50$  realizations of
disorder.}
\label{fig11}
\end{figure}

Finite size effects are also relevant in a biased lattice with long-range
correlated disorder. In Fig.~\ref{fig11} we observe a similar trend to that
found in disorder-free lattices. The main difference is related to the value of
$\tau_{\infty}$, which is slightly lower than in disorder-free lattices of the
same size.

\section{Conclusions} 

\label{conclusions}

We studied numerically the intraband exciton relaxation in a 1D disordered
lattice subjected to a linear bias of magnitude $U$. The random site potential
presents a power-law spectral density $S(k)\sim 1/k^{\alpha}$, which gives rise
to long-range correlations in site energies. Exciton transport is considered as
incoherent hoppings over the eigenstates of the static lattice, arising when the
coupling to vibrations is taken into account. The dynamics of the intraband
relaxation is monitored by means of the fluorescence decay after broadband pulse
excitation. 

Fluoresce decay in the unbiased lattice ($U = 0$) was found to be
non-exponential in both weakly ($\alpha < \alpha_c = 2$) and strongly ($\alpha >
\alpha_c = 2$) correlations limits. This time dependence reflects the existence
of many decay channels in the system with different decay times. In other words,
the wave functions vary from state to state and consequently the distribution of
the oscillator strengths is rather broad. We found a slowdown of the
fluorescence decay when $\alpha > \alpha_c$, which we relate to the peculiar
level structure of level in this case: there exist dark states below  those
states with high oscillator strength located deep in the band (like that labeled
$3$ in Fig.~\ref{fig01}).

At moderate bias the absorption spectrum broadens as compared to the unbiased
case, but no signatures of the WSL are found. The structureless spectrum is
accompanied by a non-exponential decay of the fluorescence. However, on further
increasing the magnitude of the bias, a periodic pattern is found to build up at
the center of the (already wide) absorption band when $\alpha>\alpha_c$. Its
period is equal to $U$, as for the WSL in an ideal lattice, and independent of
the system size $N$. Simultaneously, the fluorescence decay is described
approximately by a single exponential. Finally, when the parameter $\sqrt{UJ}$
exceeds the magnitude of disorder $\sigma$, the WSL in the absorption spectrum
and the exponential decay of the fluorescence appear, irrespective of the value
of the correlation exponent $\alpha$. The single exponential decay of the
fluorescence is a direct consequence of the localization by the bias, since
all states carries approximately the same value of the oscillator strength.
Deviations from perfect exponential decay were related not only to stochastic
fluctuations of the disorder but mainly to finite size effects.

\acknowledgments

The authors thank V.\ A.\ Malyshev for helpful conversations. This work was
supported by MEC (Project MOSAICO).

\end{document}